# Human Computation and Convergence

Pietro Michelucci

Human Computation Institute

**Abstract** Humans are the most effective integrators and producers of information, directly and through the use of information-processing inventions. As these inventions become increasingly sophisticated, the substantive role of humans in processing information will tend toward capabilities that derive from our most complex cognitive processes, e.g., abstraction, creativity, and applied world knowledge. Through the advancement of *human computation* - methods that leverage the respective strengths of humans and machines in distributed information-processing systems - formerly discrete processes will combine synergistically into increasingly integrated and complex information processing systems. These new, collective systems will exhibit an unprecedented degree of predictive accuracy in modeling physical and techno-social processes, and may ultimately coalesce into a single unified predictive organism, with the capacity to address societies most wicked problems and achieve planetary homeostasis.

## The First Information Processor

One could argue that information processing on Earth, and possibly in the Universe, began with the emergence of bacterial life three billion years ago(Brasier, McLoughlin, Green, & Wacey, 2006). Autopoietic life processes are fundamental to the notion of information processing(Moses, Flanagan, Letendre, & Fricke, 2013). Reproduction itself requires the ability to process and execute a set of abstract instructions encoded in DNA. Furthermore, organismal survival depends upon the ability to respond appropriately to environmental conditions, which involves sensation, response selection, and consequent action. Even archaic life forms, such as bacteria, possess these information processing capabilities, exhibiting genetic reproduction as well as movement in response to chemicals (*chemotaxis*), light (*phototaxis*) and magnetic fields (*magnetotaxis*)(Martin & Gordon, 2001).

But how does such information processing occur? The entire Behaviorist movement of the mid-20th century was agnostic to that question, treating the information-processing apparatus of an organism as a "black box", in which one would be concerned only with the environmental input and behavioral output of a system. Eventually, people who wanted to create intelligent machines came along and thought it might be useful to find out what was actually going on inside nature's black box. This led to the birth of information processing psychology and the subsequent Cognitive revolution(Miller, 2003), which gave rise to present day Cognitive Science.

Cognitive Scientists are interested in understanding *how* an environmental stimulus results in a particular response within an organism. Most explanatory models arising from this approach tend to characterize the propagation of information within an organism – in other words, an organism's *internal communication*. More specifically, such models posit that information about an organism's environment is somehow encoded and transmitted to the part of the organism that can activate a suitable response, often with mediating decisional processes. Though bacteria accomplish this via an elaborate chemosensory signaling pathway(Blair, 1995), most animals today possess complex networks of specialized cells, called neurons, that serve this purpose. Recent findings, however, suggest that even prior to the evolutionary emergence of neurons, bioelectric signaling occurred among somatic cells, a phenomenon still present in zebrafish(Perathoner et al., 2014). Thus, it would appear that information processing originated and co-evolved with life.

This raises an interesting question: What accounts for this co-evolution? In other words, what are the adaptive benefits of increasingly sophisticated information processing capabilities? An answer may be embedded in the following information-theoretic tenet: *in order for something to be informative it must be unexpected*(Desurvire, 2009). Expectations, of course, derive from predictions. But where do such predictions originate? In nature, complex organisms, such as mammals, rely heavily on prediction for survival. Anticipating, for example, the attack behavior of a predator could improve evasion. Phylogenetic and ontogenetic processes endow mammals with basic predictive models of their environment. These causal models are then dynamically fine-tuned over the life of the organism via learning. Such learning occurs when an organism experiences an unexpected outcome, which then *informs* the modification of the predictive model (see Figure 1). When successful, learning results in better predictions(Vigo, 2012). Thus, the ability to develop more complex and comprehensive predictive models of the world can lead to more accurate survival-based predictions under a wider variety of conditions and circumstances. Indeed, cumulative learning is sufficiently critical to the success of an organism that an extensive and growing body of Machine Learning research exists to emulate this capability in autonomous agent-based systems(Michelucci & Oblinger, 2010).

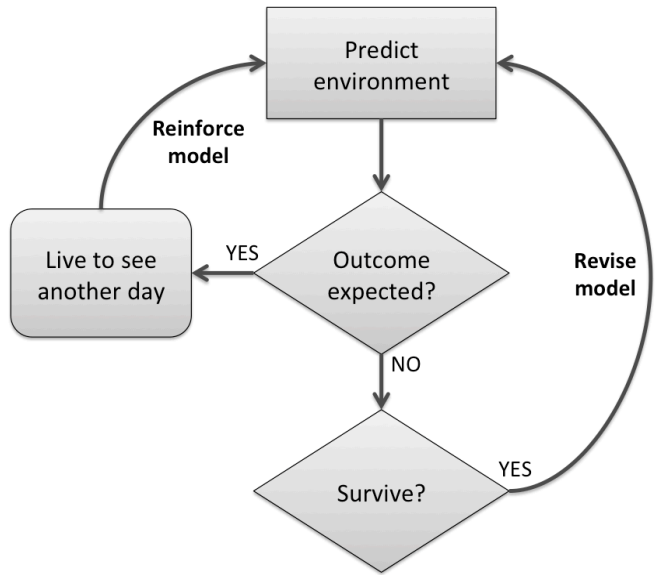

**Figure 1: Experiential learning over the course of an organism's life continually refines its predictive model of the environment.**

For most people today, the term "information processing" conjures notions of computers and network infrastructure. However, as one considers the future relationship of humans and automated systems it may be useful to reflect on the biological origins of information processing described in this section. This provides a basis for regarding humans not just as end-users of computing systems, but also as supercomputers in our own right. Such a perspective liberates us to imagine more easily what might be possible in a future world that is composed of interacting biological and artificial information processing systems, especially as those lines become increasingly blurred.

## Today's Most Advanced Information Processor

The neural circuitry of simple metazoans may limit learning to first order cause-and-effect relationships. Higher mammals, such as humans, however, exhibit working knowledge of much more complex relationships. Despite technological advancement, the human brain is widely believed to be the most sophisticated information processor in the universe, consisting of approximately $10^{11}$ neurons communicating in parallel over a network of $10^{15}$ dendritic connections(Sporns, Tononi, & Kötter, 2005; Larson-Prior, 2013; Michelucci, 2013b). However, it is not just the raw computational power of the brain that endows it with capabilities that elude our best machine-based systems, but rather its complex cognitive architecture. This neurological software, which guides the flow of information in the brain, endows humans with unique abilities such as creativity, abstraction, intuition, anomaly detection, and analytic problem solving, all in service of the most advanced predictive modeling capability known.

Thus far, biological evolution has endowed humans with the intelligence needed for survival, including the invention of powerful technologies. However, some of these inventions have led to intractable societal problems (e.g., climate change, pandemic disease, geopolitical conflict, etc.), the solutions of which exceed the reach of individual human cognitive abilities. These "wicked problems" have no specific formulation, as each problem characterization depends upon a specific solution approach, which exists among an unknown set of possible approaches(Rittel & Webber, 1973). To further complicate matters, there is no definitive endpoint; candidate solutions must be dynamic, adaptive, and ongoing. Moreover, wicked problems are multifaceted, often involving multiple systems, such that a solution that benefits one system (e.g., an ecological solution) may have repercussions to another system (e.g., financial markets).

To address these wicked problems and manage the survival risks they pose, we need an intellectual faculty more advanced than human intelligence. That is, we require more sophisticated functional and predictive models of the world and a capacity to employ those models to generate tenable and durable solutions. Because the urgency of this need exceeds the time scale of biological evolution, a viable solution will likely require technological innovation.

One such option would be to build super-intelligent machines that can be used to solve our wicked problems for us. Some believe that creating human-like cognition in machines, often referred to as Artificial General Intelligence (AGI), will happen in three steps: 1) map the entire human "connectome", the neural circuitry of the brain, 2) replicate that circuitry using computer hardware and software, and 3) wait until microprocessor speeds exceed the processing speed of the human brain. The culmination of these steps has been referred to as the "Technological Singularity"("Technological singularity," 2014), which refers to the moment at which AGI exceeds human intelligence, and beyond which we cannot anticipate the impact on civilization (hence the metaphor to the "singularity" construct in Physics, which involves a "point of no return").

This approach may not be tenable for two reasons. First, there is wide disagreement among Artificial Intelligence experts about when the Technological Singularity might occur(Armstrong & Sotala, 2015), ranging from just a few years to beyond the year 2100. Thus, such an event may not be soon enough to mitigate deleterious global processes before they become irreversible. A second reason is that most predictions about the Technological Singularity are predicated on replicating the human connectome in-silico. However, recent evidence(Hansson & Rönnbäck, 2003)(Tang et al., 2014) detailing the role of non-neuronal cells in brain signaling is emblematic of a fundamental issue with this approach: that a connectomic model of the human brain, which maps the interconnection of existing neurons, along with a functional model of neurotransmission, may be inadequate for explaining cognitive function. Indeed, we are far from having a comprehensive understanding of the biochemical mechanisms that govern brain development and activity. For this reason, developing human-like cognitive abilities in machines will not depend solely upon processing speed or artificial replication of the connectome, but will also rely upon the realization of a complete and accurate dynamic, mechanistic model of human cognitive processes(Michelucci, 2013a).

## Human Computation

An alternative approach to building superior intelligence, which may have greater near-term promise, involves directly leveraging human cognition within a distributed information-processing network. As previously described, there are cognitive abilities that remain solely within the purview of humans despite progress in artificial intelligence research. However, there are also tasks much better entrusted to machines than humans, such as counting, calculating mathematical and statistical formulas, and keeping track of events and outcomes. Indeed, the information processing architectures and resultant capabilities of machines and humans (see Figure 2) exhibit a striking complementarity that suggests an opportunity for fruitful collaboration.

| Machines | Humans |
|---|---|
| • Counting | • Inference |
| • Precision | • Visual Perception |
| • Objectivity | • Linguistic Ability |
| • Calculation | • Abstraction |
| • Persistent Storage | • World Knowledge |
| • Data Integrity | • Sociocultural Awareness |
| • Process Execution | • Creativity |

Figure 2: Complementary information processing strengths of machines and humans.

What if it were possible to engineer systems that combine the respective strengths of machines and humans toward unprecedented information processing capabilities? Would this endow humanity with a problem-solving capacity sufficient to address present and future societal challenges? Over the past decade, a scientific community has emerged to explore the transformative potential of directly employing human cognition within larger computational systems, giving rise to a field of study called "human computation". Human Computation (HC) has been defined generally as the design and analysis of information processing systems in which humans participate as computational elements(Michelucci, 2013c). In this definition, the notion of "computation" encompasses the full spectrum of processes that might be applied to the transformation, synthesis, and interpretation of data.

HC systems can be classified at a high level as either naturally emergent or deliberately engineered(Michelucci, 2013c). As previously described, information processing is integral to most natural human behavior. When large collections of such behaviors are accessible via the technosocial infrastructure, it becomes possible to extract useful information from such data. In this vein, human computation can manifest as an *emergent* phenomenon. One example of this is financial markets, the aggregate behaviors of which sometimes can be used to predict world events better than individuals, exhibiting the "wisdom of crowds" effect. Another example of emergent HC involves an analytic approach called "query based syndromic surveillance", which examines trends in online search behavior. Incorporating prevalence measures of semantically relevant Google search terms (e.g., Influenza symptom descriptions, Influenza terms, Influenza complications, etc.) has been shown to enhance Influenza forecasting models(Dugas et al., 2013)(Ginsberg et al., 2009).

Most HC research today, however, focuses on how to *engineer* goal-directed HC systems, which leverage human cognitive capabilities that still exceed the best automated methods. The HC community has begun to recognize and catalog classes of HC systems toward understanding the success precursors for those systems and improving the repeatability of attendant methods(Greene, 2013). To develop a more concrete sense of the use and impact of human computation today, it is worthwhile to examine some of the more prevalent classes of HC systems along with notable exemplars.

As an early success case, reCAPTCHA has become a canonical example of human computation, and in particular, of a *crowdsourced micro-tasking* system. By embedding reCAPTCHA into their websites, Internet entities can distinguish between legitimate human users and mal-intentioned web-crawlers. This is accomplished by making access to site content contingent on the user entering words seen in distorted text images, which today is easier for a human than a machine. In addition to providing a human verification service, reCAPTCHA simultaneously helps curate digital archives by including failed OCR fragments among the examples of distorted text so they can be resolved by the human responses. Thus, in the course of proving their humanness, hundreds of millions of reCAPTCHA users have unwittingly contributed to digitizing a century's worth of archival issues of the New York Times.

"Quid pro quo" systems, like reCAPTCHA(Ahn, Maurer, McMillen, Abraham, & Blum, 2008), compel participation by offering some service (e.g., access to Web content) in exchange for human computation labor. Another system manifesting such reciprocity is DuoLingo(Garcia, 2013), which provides foreign language instruction during which the users' online learning behaviors contribute directly to text translation services. Thus, DuoLingo students have, through their foreign language studies, collectively translated Wikipedia articles from one language into another. The striking observation that Duo-Lingo implicitly trains its own labor market is suggestive of the future potential of HC to become a disruptive economic force.

In contrast to quid-pro-quo HC systems, crowdsourcing marketplaces (e.g., Amazon Mechanical Turk, CrowdFlower, etc.) provide monetary compensation as an extrinsic motivator. Microtasks, such as tagging photos, completing surveys, or editing documents is outsourced to a community of active "crowd workers". Remuneration is often predicated on a minimum level of performance and paid in proportion to quantitative contribution metrics, as determined by the micro-task stakeholder, who is typically referred to as the "requestor". Such crowdsourcing marketplaces have been used historically for market research but are also being used increasingly by scientists to collect data, a practice sometimes referred to as *Cyberscience*(G. Newman, 2014).

Historically, checks and balances in crowdsourcing marketplaces have catered primarily to the requesters. However, in response to concerns about the unscrupulous practices of some requestors, reputation systems such as TurkOpticon have emerged to help crowd workers themselves make more informed choices about which projects and project requestors they choose to engage.

*Citizen Science* is an application of human computation that refers broadly to public participation in the scientific process (G. Newman, 2014), typically in an online context, but not always (e.g., The Audubon Christmas Bird Count(Ullrich Barcus, n.d.) and The St. Louis Baby Tooth Survey(Early, n.d.)). An emerging differentiator between Citizen Science and Cyberscience is that the former seems to necessarily involve an educational aspect, such as participant understanding of the scientific goals and cognizance of how participation contributes to those goals. Irrespective of these potential educational benefits, project instigators are often drawn to the prospect of the tremendous research acceleration that can result from harnessing the power of the crowd. Indeed, most citizen science projects involve crowdsourced microtasking that involves some aspect of data curation that lends itself to human cognition. Moreover, to retain participation, many such projects include "ludic" or game-like elements(Krause, 2013).

A pioneering example of online citizen science is the stardust@home project, which began in 2006, in which 30,000 participants used a virtual microscope to analyze millions of digitized images of aerogel to detect nano-scale cosmic dust particles retrieved from the tail of comet Wild 2. To be successful, participants had to learn to use a virtual focuser to visually distinguish particle trails from prevalent aerogel inclusions. Ultimately, the project discovered seven particles deemed likely to be extra-solar particle, the composition of which has suggested revision of existing models of cosmology. All 30,000 par-ticipants were included as co-authors on the paper reporting these results in the journal *Science*(Westphal et al., 2014).

Citizen science is being used increasingly in the medical research field (e.g., fold.it, Phylo, WeCureALZ, etc.), particularly by leveraging the analytic ability of humans to exclude avenues of discovery that are unlikely to be fruitful. Since automated solution search methods tend to be exhaustive, they can be computationally intensive and even intractable for certain classes (NP-hard) of problems. Human-based approaches, however, tend to employ world knowledge, abstract reasoning, transfer learning, and intuition to identify a handful of fruitful avenues worthy of further exploration. The fold.it project has gambled successfully on these human abilities, by recasting a complex biomedical research problem into a 3D folding game accessible to non-specialists. Even though thousands of participants provide puzzle solutions, it turns out that only the very best human-based solutions have exceeded machine-generated solutions, and thereby enable new research discoveries. Thus, in contrast to the stardust@home project, which employs a "divide-and-conquer" approach to finding needles of comet dust within a proverbial haystack of digital imagery, the fold.it project uses a "winner-takes-all" model to determine optimal protein molecule configurations. To date, fold.it puzzle solutions have been instrumental in advancing our understanding of the Simian Immunodeficiency Virus (SIV), a close relative of HIV. Fold.it is now setting its sights on proteins implicated in the Ebola virus(Long, n.d.).

More recently, architects of citizen science systems have taken a greater interest in collaborative discovery, perhaps inspired by the community-based discovery of a new type of "green pea" galaxy due to the inclusion of an online social forum in the GalaxyZoo project(Cardamone et al., 2009). The Phylo project has created Open-Phylo to allow any disease researchers to upload genetic sequences for crowd-based analysis through the Phylo platform. Phylo's creator, Jérôme Waldispühl, views this new portal as a potential opportunity to use the solution from one Phylo participant as the starting point for another participant toward collaborative solution development.

ReCAPTCHA has been described above as a quid-pro-quo system, in which a trade relationship exists between each crowdworker and the project stakeholder. Over the past few years, however, a different sort of reciprocity has fueled a class of human computation platforms, in which the community of participants is itself a stakeholder and shares goals with individual participants. In these *virtuous ecosystems*, individual participants contribute information, which is then combined with other users' information and shared back as an aggregate. The platform serves this aggregated information in a form that is easily understood and tailored to the specific needs of each individual, leading to modified behaviors in the real world and often to improved individual results. In some cases, participants will continue to report back new results to the system and perpetuate this virtuous circle. Note that "virtue" in this context, refers not to the application domain but rather to the virtue of contributing to a greater whole, which in turn benefits individual participants in ways that encourage further contributions.

One clever and effective example of such a virtuous ecosystem is enabled by an online platform called "Trapster". In Trapster, participants are vehicle drivers who click a button on a smartphone app whenever they see a speed trap along the road. Using georegistration methods, a central server records the location and time of the speed trap observation, which then contributes to a mashup that reports back to Trapster users the location and recency of any speed traps in their immediate path so they can avoid citations. Ironically, Trapster has resulted overall in improved speed limit compliance, which suggests serendipitous benefits to unintended stakeholders, such as law enforcement organizations and perhaps local pedestrians.

Other noteworthy examples of virtuous ecosystems include PatientsLikeMe and the YardMap project. PatientsLikeMe enables a community of people with certain medical conditions to report regularly about their current symptoms and treatments. The aggregated data is used to provide intuitive graphs and charts showing treatment efficacy tailored to the specific clustering of diseases and symptoms of each patient, based on what has been working for other patients with similar disease and clinical presentation profiles. Ancillary stakeholders are medical research organizations that purchase the anonymized data to develop more effective treatments.

The Cornell Lab of Ornithology's YardMap Project allows online participants to outline any site on a shared aerial map, producing detail maps of spaces ranging from yards or roof garden to a parks or corporate campus, and identifying habitat types, individual objects and practices (e.g., plants, solar roof panels, composting). Users can ask questions about their plants, habitats, practices and more knowledgeable users can then view the annotated yard maps and provide environmental conservation advice in the integrated social network to facilitate learning and help newcomers to become resources for future users(Dickinson & Crain, 2014). Communities also use the platform to work together toward shared goals, such as increasing the percentage of native flora in a neighborhood or enhancing pollinator habitat. YardMap thus generates useful conservation data that can be tied to the Lab's bird monitoring efforts, while also serving as a powerful platform for leveraging online social dynamics to influence real world behaviors that are aligned with users' goals(Dickinson, Crain, Reeve, & Schuldt, 2013).

Though such virtuous ecosystems manifest a karmic benefit to contributors, it is not difficult to imagine nefarious applications. To better understand such risks, this topic was explored by a breakout team at a recent workshop((multiple authors), 2015), who considered applying methods such as cascading social networks to amplify seeds of disinformation(McDonald et al., 2014).

The advent of the Internet removed physical barriers to human collaboration - a brick and mortar meeting room can hold 30 people, but a virtual room can hold 30 million peo-ple. This meant that information processing workflows involving contributions from people, machines, or both could be automated to enable collective innovation and problem solving at unprecedented scales. For example, applying machine-based aggregation methods to such distributed information processing enables a "wisdom of the crowds" effect(Surowiecki, 2005), in which the collective answer to a problem exceeds the best individual answer. This effect has been demonstrated from problems ranging from the recollection of order information (Steyvers, Lee, Miller, & Hemmer, 2009) (e.g., in what or-order were these 20 books published) to solving the traveling salesman problem(Yi, Steyvers, Lee, & Dry, 2012) (i.e., given a roadmap, what is the shortest path from point A to point B?). It turns out that the specific algorithms and underlying theory used for combining human inputs is critical to the success of these methods(Yi et al., 2012).

In contrast to wisdom of crowd approaches, which tend to be non-social and rely on algorithmic approaches to combining human answers, other approaches to collective intelligence seek to enhance human collaboration. Indeed, Francis Heylighen, of the Global Brain Institute, defines collective intelligence as "a group's ability to solve problems and the process by which this occurs"(Michelucci, 2013c). Anita Woolley and her colleagues have further explored this notion, by inventing a "Group IQ" metric as a way to investigate factors that influence the problem solving ability of small groups(Woolley, Chabris, Pentland, Hashmi, & Malone, 2010). Woolley, et al., made the noteworthy discovery that the single most important success factor in group-based problem solving is social intelligence rather than individual IQ. In other words, the individual problem solving ability of group members is less relevant to problem solving success than the ability of the group to work together.

NASA's Center of Excellence for Collaborative Innovation (CoECI) has been a pioneer in driving collective innovation, the successes of which have prompted interest in this approach by other U.S. federal agencies, including the Department of Homeland Security and the National Institutes of Health. In NASA's tournament-style challenges, a problem description is provided to a solver community in which prizes are simply awarded to the best solutions. In one noteworthy case, a challenge was posed to improve NASA's ability to forecast solar proton events (resulting from solar flares), which pose radiation hazards to spacecraft and astronauts. Previous NASA and academic efforts could predict such events one to two hours in advance. The winning solution, provided by a retired radio engineer, could predict such events eight hours in advance with 85% accuracy(Davis & Richard, 2010).

Though not inherently collaborative, such tournament-style competitions can lead to collaboration when mutual awareness prompts teams to combine approaches in order to be more competitive. Such spontaneous collaboration occurred during the 2009 Netflix Prize challenge, which sought a better algorithm for predicting user film recommendations based on prior recommendations. A $50K prize was offered to the best solution, but to obtain the million-dollar grand prize, the winning solution had to be at least 10%

better than the existing "Cinematch" recommender system. Two teams were close, but neither exceeded the 10% improvement threshold required for the grand prize, so they decided to combine their solutions, which resulted in a 10.09% improvement and a substantially greater monetary award("Netflix Prize," 2015) even when splitting the prize.

Internally, NASA employs a more collective (and less competitive) form of innovation, using third-party platforms such as InnoCentive@work, in which problem-solving is driven by an online collaborative workflow process that elicits problem definition, facilitates collaborative discussion, and streamlines solution evaluation. More sophisticated distributed problem solving approaches are also emerging, such as the ePluribus problem solver, which employs stigmergic (i.e. indirect coordination, such as the scent trails left by ants) methods to enable users to collaboratively blaze solution paths between an initial state and goal state. Within this path-finding context, ePluribus also facilitates problem decomposition into manageable components that can be addressed by users asynchronously, and then fuses the sub-solutions from the diverse feedback of many problems solvers into collective solution paths(Greene & Young, 2013). Such an approach is potentially scalable to millions of users.

Some might view the ongoing activities of the scientific community as a form of collective intelligence. Indeed, execution of the scientific process and communication of empirical findings is becoming increasingly automated via collaboration platforms, public engagement through citizen science, online journal management systems, experimental metadata capture(Shreejoy et al., 2013), and candidate approaches for connecting these islands of automation within a coalescent framework for collective science(Thierry Buecheler, 2010). Still, today science tends be a sequential process in which a single research team produces hypotheses, methods, findings, and conclusions, which are all published after the study concludes.

One might imagine, however, a platform for collaborative science in which scientific communication coevolves with research and contributions can be inserted by anyone at any stage of the process. Adopting versioning techniques from the software development field, such as branching processes, would permit a single hypothesis to be tested in multiple ways, and for the resultant data to itself be subjected to various analyses and conclusions, each giving rise to "version 2" studies, in which new hypotheses are motivated by the version 1 study results, all within a common framework of traceable activities, data, and results. This new form of open research would better play to individual strengths, improve transparency, permit instantaneous knowledge transfer, and facilitate replication. On the other hand, it would pose a new set of challenges, such as how to implement credit assignment. For example, person A would have to be credited for providing a hypothesis, person B for defining methods, person C for improving upon person B's methods, person D for statistical analysis, and so on. Nonetheless, connecting such a collaborative research platform to various crowd-powered systems for data collection and curation could lead to a culture of pervasive participatory science and markedly accelerated scientific advancement.

The collective intelligence approaches discussed thus far entail specific, well-defined methods or processes to achieve certain outcomes. Indeed, our modus operandi for collective intelligence has been, essentially, to insert humans at various stages within an algorithmic process. Human cognition, however, which represents our most sophisticated and capable model of intelligence, seems to manifest multiple levels of abstraction and emergent properties, such as consciousness, arising from its complexity. Indeed, the way we think is decidedly not algorithmic. How, then, might we achieve collective intelligence more closely modeled on human cognition and what might we hope to gain from it?

Marvin Minsky, a renowned cognitive scientist and artificial intelligence pioneer, published a popular book in 1988 called The Society of Mind(Minsky, 1988), based on a radical new theory co-developed with Seymour Papert that the human mind is functionally composed of myriad special-purpose agents, and that it is the interaction of these agents that results in high level cognition. Following Minsky's work, others came along and made specific commitments about the interactions of such agents in various attempts to functionally explain human cognition (see (Chong, Tan, & Ng, 2009)). These cognitive theories, as well as other more biologically inspired approaches, were implemented in software as *cognitive architectures*, which purport to functionally replicate goals, memory, learning, reasoning, and other human cognitive faculties. Today, no single cognitive architecture provides a comprehensive working model of human cognition, but this line of research continues to bring us closer to simulating human intelligence.

This notion of cognitive architecture provides a context for thinking about how one might engineer humanlike collective intelligence. Consider a promising cognitive architecture as a starting point, and for each agent-based system or specialized process within that architecture, imagine substituting a human or human collective that was appropriately specialized by design. Might it be possible in this way to realize a fully cognitive system that does not merely exhibit crowd-based wisdom on a specialized task, but rather manifests collective goals, via executive functions that gather, process, and act on information at massive scales? Though such a *hybrid cognitive architecture* may be a plausible approach for achieving humanlike collective intelligence, it is conceivable that achieving superhuman collective intelligence will require entirely new cognitive architectures that maximize synergistic effects among all agents, whether human or artificial.

With the societal introduction of collective intelligence comes the risk of collective psychopathology. Human mental illness (e.g., neurosis and psychosis) can often be recast as information processing dysfunction(Endres, 2011), which could also occur in collective information processing systems(Blumberg & Michelucci, 2013). The potential interplay of mental illness in individual human contributors and systemic psychopathology in the human collective remains to be studied. Thus, as we become increasingly depend-ent upon systems exhibiting collective intelligence, it would behoove us to retain awareness of this potential risk and consider mitigation strategies.

As discussed, collective intelligence research tends to focus on understanding how intelligence emerges from the interaction of many individuals and how to use that knowledge to engineer systems that can solve problems more effectively than their constituent human or machine contributors can solve individually. Although such collective intelligence systems may be used to process real world data and may themselves be composed of physical systems (e.g., humans and machines), they do not necessarily interact directly with the physical world.

Nonetheless, human-based distributed sensing is an evolving research area in its own right, including such approaches as *collective sensing*, which involves applying statistical methods to aggregated social network data, *people as sensors*, which involves humans in the wild contributing their subjective perceptual experiences via mobile devices, and *participatory sensing*, in which humans use portable sensor technology to acquire objective data from the world(Resch, 2013). One clever implementation of participatory sensing is StreetBump, a smartphone app that runs in the background using the device's built-in accelerometers to sense potholes in the road when people drive over them. When a "bump" is sensed, it is reported with GPS data to a central server where it is added to a pothole map that helps inform future roadwork(Carrera, Guerin, & Thorp, 2013). In this example, participatory sensing capitalizes on existing patterns of human behavior and, therefore, does not require any new human activity other than downloading and running the app.

The other two distributed sensing methods described above, *people as sensors* and *collective sensing*, have been useful in crisis relief efforts(Meier, 2013). Ushahidi's CrowdMap platform enables the ad hoc creation of a geospatial mash-up, in which users can contribute subjective, locale-specific information to an evolving map. This *people as sensors* platform was employed during the 2010 Haiti Earthquake to enable anyone to text their needs to a 4-digit code and have that request automatically geolocated on a map. In the wake of Typhoon Pablo, which devastated the Philippines in 2012, *collective sensing* was employed for damage assessment, resulting in the collection of 20,000 relevant tweets. This, however, posed a new problem, which was how to derive actionable information from that social network data. Interestingly, the UN task force decided to use crowdsourcing to analyze the collective sensing data. CrowdFlower, a firm specializing in crowdsourced microtasking, distributes the analysis of those tweets to thousands of crowd workers, who identified any links leading to photos or videos and then assessed the imagery for evidence of damage. Within twelve hours, the task force was able to produce a georeferenced damage assessment map with linked imagery, all due to collective sensing and crowdsourced social network analysis.

In addition to amplifying awareness through distributed sensing, human computation also affords an opportunity to be a more effective in the world via coordinated action.

Swarm theory demonstrates how local coordination can result in global emergent behaviors (e.g., the V-shaped formation of a migrating gaggle). However, the opportunity for centralized coordination afforded by mediating technology could be used to share locally relevant information with individual actors to improve their efficacy, and also to implement more complex activities. WikiProject is an automated tool that seeks to coordinate the online activities of Wikipedia contributors by leveraging the availability of user activity information. By calling attention to the activities of other users who share the same goals, WikiProject attempts to encourage contribution and reduce redundant effort, though perhaps with limited efficacy(Riehle, n.d.).

While collective intelligence, distributed sensing, and coordinated action, when implemented successfully, are each potentially transformative in their own right, the prospect of combining those capabilities within a unified system suggests a tantalizing opportunity to build a distributed organism that manifests collective agency in the world. Such a "superorganism" would exhibit pervasive awareness through its distributed sensory faculties, reason with an unprecedented degree of predictive accuracy, and implement complex, multi-actor behaviors. This model has evolutionary precedents among the eusocial insect species, which derive survival advantages through locally cooperative, globally emergent collective behaviors. Indeed, a recent comparative analysis of these insect behaviors to open source software development has provided inspiration for new human computation methods(Pavlic & Pratt, 2013).

The effects of organismic computing were investigated in an online hide-and-seek game to measure the impact of shared sensing, collective reasoning, and coordinated action on group efficacy(Michelucci, 2013b). Employing a novel technique, called simulated augmented reality, made it possible to test collaborative enhancements that have not yet been engineered. In contrast to virtual reality, which simulates the world in a computer-generated environment, augmented reality superimposes virtual elements onto the real world via pass-through displays. Because augmented reality technology today is still relatively immature, simulating augmented reality in a virtual environment serves as a useful proxy.

The overall benefit of organismic computing on group performance was supported by the study results, as was the central prediction that larger groups would benefit more from organismic computing than smaller groups. These early findings hint at the transformative capabilities that might arise due to organismic computing implemented on a massive scale. Organizational units at every scale, from businesses(Brambilla & Fraternali, 2013) to governments, from societies to countries, and even humanity in its entirety could benefit from the synergy, efficiencies, and awareness that might arise through these methods.

# Coevolution, Convergence, and Emergence

Organismic Computing, as discussed in the previous section, describes a technological trend toward the interconnectedness of humans across space. To see the complete picture, however, requires us to step back and consider the impact of interconnectedness over time. Smaldino and Richerson describe *cumulative culture* as "learned information and behaviors [that] are reliably transmitted and improved upon" from one human generation to the next(Smaldino & Richerson, 2013), pointing out that no single human being knows how to build a modern computer from scratch. Thus, cumulative culture serves as long-term memory for humanity, ensuring that each new generation does not have to re-invent the wheel (or the computer, for that matter). But where does that memory reside?

Along the way, technology itself has improved our ability for cumulative cultural evolution. Though generations long passed may have propagated wisdom and knowledge through storytelling and mentorship, the advent of written communication, and methods to preserve such writing, extended the state space of human knowledge from our minds to the external, physical world. Eventually, we developed abstract symbolic methods for specifying a process, such as a mathematical formula, and even built crude mechanical devices for executing such processes, all the time, carefully recording our methods in writing so that future generations could build on that knowledge. And, indeed, as we became more advanced in our ability to harness the power of nature, we could use that knowledge to replicate these computational processes by routing electrons through semi-conductors, and even using them to store information. This knowledge has enabled us, through increasingly sophisticated methods, to extend information processing and storage to the external world and increase the fluidity with which such information passes between humans and the physical universe. Thus, the very technology that has been advanced through intergenerational collaboration has been further streamlining that collaboration, thereby accelerating the co-evolution of technology and human society in a self-perpetuating process.

The foregoing analysis suggests that the convergence of human systems and technology has led to increasing connectedness among humans over both time and space, and that this has been enabled by technology implemented in physical systems. In order to better frame the potential implications of increasingly convergent human systems and technology, it might be useful to first consider a simpler coupling. Imagine a student who is accustomed to adding 3-digit numbers in her head, which is not too difficult because most humans can retain that amount of information in short-term memory. Now she is given a more difficult arithmetic problem, which involves adding two 10-digit numbers. Because it is difficult to remember so many digits, the student decides to use paper and pencil to perform the computation. This raises an interesting question: with the paper serving as a state space (i.e., a working memory) for solving the problem instead of activation patterns in her brain, is part of the thinking now happening outside of her brain? Clark and Chalmers have suggested that *epistemic actions*, that is, the employment of external physical constructs to augment cognition, constitutes an extension of the human mind(Clark & Chalmers, 1998). This view considers the human organism and facilitative physical constructs as a coupled cognitive system.

Extending this notion of a coupled cognitive system to include all human systems at different scales, and all physical systems that assist humans in processing, sharing, and storing information suggests a new framework for understanding and anticipating the convergence of human systems and technology.

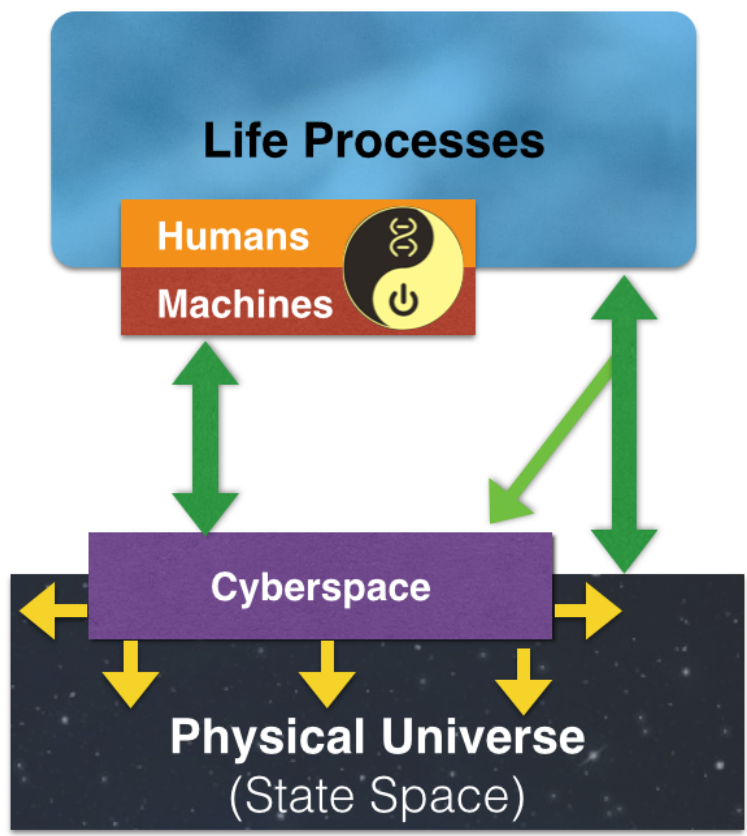

Figure 3: A cognitive model of the universe.

In this framework (see Figure 3), the physical universe is viewed as a canvas upon which life processes encode knowledge. Humans are viewed as information processors, aggregators, and actors that serve as intermediaries between the physical state space and the virtual cyberspace, encoding knowledge about the former in the latter, and using that knowledge, in turn, to manipulate the former. At its essence, cyberspace can be thought of as an economical representation of information, which subsists upon a physical infrastructure that provides communication, data storage, and algorithmic computing. In practice, however, cyberspace is the medium within which humans represent, share, and build an understanding of the universe. Thus, as human interaction with cyberspace becomes more pervasive, frequent, and varied, the resultant coupled cognitive system develops a more faithful model of itself.

Today, our shared model of the universe is largely piecemeal, spread across expert minds and specialized communities. To make matters worse, we are flooded daily by petabytes of physical, genomic, social, economic, and other varieties of data. Thus, except for domain-specific computational simulations that exist in isolation, the majority of our knowledge is represented statically, encoded in human language and our data is distributed across heterogeneous repositories. Fortunately, human computation systems such as Wikipedia are improving this situation via collaborative knowledge capture. Furthermore, infrastructure that supports human computation, such as the Semantic Web, promises to enable new human-based discoveries and more usable encodings by linking isolated datasets(Simperl, Acosta, & Flöck, 2013). Over time, we can expect increasingly sophisticated human computation methods to help us leverage our cumulative culture and growing pile of data into cohesive, working knowledge of the world – that is, predictively accurate, dynamic models of the interactions that occur among physical systems including life forms. In fact, these models would be self-reflective, as human contributors would also be modeling themselves and their relationships with other human and physical systems, including the infrastructure upon which the models are built. Ultimately, we might imagine this culminating in a system that is best able to help us predictably influence the world in desirables ways.

The organismic view of information processing described earlier suggests that survival requires an organism to employ accurate predictive causal models of itself and its environment. When this view is applied, in the large, to the coupled cognitive system of life processes interacting with the physical universe, such predictive models become self-perpetuating meta-models. In other words, when life forms collaborate and coalesce, as enabled by technology, to produce a more advanced predictive model of the universe, they are better able to self-adjust and engineer effective interventions that further perpetuate life and more advanced information processing systems, leading to yet better predictions. Dynamic systems theory would suggest that such a cascading process could lead to a phase shift, that is, a sudden qualitatively different pattern of organization in the life-universe system. One conceivable manifestation of this phase shift is consciousness. Integrated Information Theory measures human consciousness in terms of the causal impact of abstract information processing on subordinate constituent processes, stipulating a minimal level of connectivity among processing elements necessary for the emergence of consciousness(Tononi, 2008). Thus, a sufficient level of information processing capacity and complexity within the life-universe system may lead to universal consciousness and self-awareness. In such a tightly integrated system, one could imagine the manifestation of collective thoughts that are as transcendent to the thoughts of an individual human as human thoughts are to the simple information processing of a constituent neuron.

## Conclusion

Each year, data production increases exponentially. This trend will continue as devices, such as home appliances, become increasingly Web-enabled. The "quantified self" movement alone will result in millions of personalized, sensor-based streams of physiological data captured during exercise. However, in order for data to improve our understanding of the world, it must be informative, which requires both a goal-directed context and an information processor that can extract knowledge that has predictive value. Human computation, in some sense, represents an enabling technology for a new "Web of Agents" that will help make sense of the massive quantity of data generated by the Internet of Things. Methods such as massively distributed problem solving, citizen science, distributed analysis, and collaborative modeling, will efficiently employ human and machine agents in complementary fashion to convert big data into useful information.

The convergence of humans and technology will continue to accelerate due to emerging technologies, such as neural implants and augmented reality, that expand the purview of human experience to more viscerally incorporate cyberspace into a mixed reality. The information ecosystem will rapidly mature with both human and machine based consumers and producers of information, resulting in coalescence across domains (e.g., social, financial, political) and scientific disciplines. Massively collaborative methods will dynamically embed values that translate into policy and evolving models of governance. Increased automation, such as driverless cars(J. Newman, 2014), will gradually compel humans to occupy a narrower but more satisfying set of labor categories that play to uniquely human cognitive abilities, such as abstraction, intuition, creativity, and discovery such that contentment occurs even "below the API"(see Reinhardt 2015).

Ultimately, we will find new ways to incorporate non-human life processes into our collective models (e.g., mammals, bacteria, viruses). We will develop interfaces at the nano-scale that support the use of microbe-based sensing and information processing. We will develop new ways of propagating and storing information. Over time the physical substrate will become increasingly economical and indelible(see Hornyak 2012) as a state-space for accumulated knowledge. Our models will include meta-models describing their own function. Ultimately, the life systems that emerged due to the specific combination of quantum particles and forces that define our universe will be able to collectively describe and understand the mechanisms of their own emergence and through that understanding create, new increasingly complex systems, which will in turn produce ever higher-level explanatory meta-models. The affordances of such runaway complexity are difficult to anticipate, but it is conceivable that such a system will become increasingly aware, intelligent, and influential.

20Chong, H.-Q., Tan, A.-H., & Ng, G.-W. (2009). Integrated cognitive architectures: a survey. *Artificial Intelligence Review*, *28*(2), 103–130. doi:10.1007/s10462-009-9094-9

Clark, A., & Chalmers, D. (1998). The Extended Mind. *Analysis*, *58*(1), 7–19. doi:10.1093/analys/58.1.7

Davis, J. R., & Richard, E. E. (2010). Advancing Innovation Through Collaboration: Implementation of the NASA Space Life Sciences Strategy. Presented at the 62nd International Astronautical Congress, Cape Town, South Africa. Retrieved from http://ntrs.nasa.gov/search.jsp?R=20110014797

Desurvire, E. (2009). *Classical and Quantum Information Theory: An Introduction for the Telecom Scientist*. Cambridge University Press.

Dickinson, J. L., & Crain, R. L. (2014). Socially Networked Citizen Science and the Crowd-Sourcing of Pro-Environmental Collective Actions. In N. Agarwal, M. Lim, & R. T. Wigand (Eds.), *Online Collective Action* (pp. 133–152). Springer Vienna. Retrieved from http://link.springer.com/chapter/10.1007/978-3-7091-1340-0_8

Dickinson, J. L., Crain, R. L., Reeve, H. K., & Schuldt, J. P. (2013). Can evolutionary design of social networks make it easier to be "green"? *Trends in Ecology & Evolution*, *28*(9), 561–569. doi:10.1016/j.tree.2013.05.011

Dugas, A. F., Jalalpour, M., Gel, Y., Levin, S., Torcaso, F., Igusa, T., & Rothman, R. E. (2013). Influenza Forecasting with Google Flu Trends. *PLoS ONE*, *8*(2). doi:10.1371/journal.pone.0056176

Early, R. (n.d.). How to Stop a Nuclear Bomb: The St. Louis Baby Tooth Survey, 50 Years Later. Retrieved January 25, 2015, from http://www.stlmag.com/St-Louis-Magazine/October-2013/How-to-Stop-a-Nuclear-Bomb-The-St-Louis-Baby-Tooth-Survey-50-Years-Later/

Endres, M. J. (2011). *An Information Processing Account of Behavioral Disinhibition in Externalizing Psychopathology*. INDIANA UNIVERSITY. Retrieved from http://gradworks.umi.com/34/91/3491470.html

Garcia, I. (2013). Learning a Language for Free While Translating the Web. Does Duolingo Work? *International Journal of English Linguistics*, *3*(1). doi:10.5539/ijel.v3n1p19

Ginsberg, J., Mohebbi, M. H., Patel, R. S., Brammer, L., Smolinski, M. S., & Brilliant, L. (2009). Detecting influenza epidemics using search engine query data. *Nature*, *457*(7232), 1012–1014. doi:10.1038/nature07634

Greene, K. A. (2013). Introduction to Techniques and Modalities. In P. Michelucci (Ed.), *Handbook of Human Computation* (pp. 279–283). Springer New York. Retrieved from http://link.springer.com/chapter/10.1007/978-1-4614-8806-4_23

22

24